\documentclass[preprint,12pt]{aastex}

\shortauthors{Winn et al.~2007}
\shorttitle{Spin-Orbit Alignment in HD~147506}

\begin{document}

%
\def\ltsima{$\; \buildrel < \over \sim \;$}
\def\lsim{\lower.5ex\hbox{\ltsima}}
\def\gtsima{$\; \buildrel > \over \sim \;$}
\def\gsim{\lower.5ex\hbox{\gtsima}}
\def\lam{\lambda=0.1 \pm 10\fdg7}
                                                                                          
%

\bibliographystyle{apj}

\title{
Spin-Orbit Alignment for the Eccentric Exoplanet HD~147506b$^1$
}

\author{
Joshua N.\ Winn\altaffilmark{2},
John Asher Johnson\altaffilmark{3},
Kathryn M.~G.~Peek\altaffilmark{3},\\
Geoffrey W.\ Marcy\altaffilmark{3},
Gaspar \'{A}.\ Bakos\altaffilmark{4,5},
Keigo Enya\altaffilmark{6},
Norio Narita\altaffilmark{7},\\
Yasushi Suto\altaffilmark{7},
Edwin L.\ Turner\altaffilmark{8},
Steven S. Vogt\altaffilmark{9}
}

\altaffiltext{1}{Data presented herein were obtained at the W.M.~Keck
  Observatory, which is operated as a scientific partnership among the
  California Institute of Technology, the University of California,
  and the National Aeronautics and Space Administration, and was made
  possible by the generous financial support of the W.~M.~Keck
  Foundation.}

\altaffiltext{2}{Department of Physics, and Kavli Institute for
  Astrophysics and Space Research, Massachusetts Institute of
  Technology, Cambridge, MA 02139, USA}

\altaffiltext{3}{Department of Astronomy, University of California,
  Mail Code 3411, Berkeley, CA 94720, USA}

\altaffiltext{4}{Harvard-Smithsonian Center for Astrophysics, 60
  Garden Street, Cambridge, MA 02138, USA}

\altaffiltext{5}{Hubble Fellow}

\altaffiltext{6}{Department of Infrared Astrophysics, Institute of
  Space and Astronautical Science, Japan Aerospace Exploration Agency,
  3-1-1, Yoshinodai, Sagamihara, Kanagawa, 229-8510, Japan}

\altaffiltext{8}{Department of Physics, The University of Tokyo, Tokyo
  113-0033, Japan}

\altaffiltext{9}{Princeton University Observatory, Peyton Hall,
  Princeton, NJ 08544, USA}

\altaffiltext{10}{UCO/Lick Observatory, University of California at
  Santa Cruz, Santa Cruz CA USA 95064}

\begin{abstract}

  The short-period exoplanet HD~147506b (also known as HAT-P-2b) has
  an eccentric orbit, raising the possibility that it migrated through
  planet-planet scattering or Kozai oscillations accompanied by tidal
  dissipation. Either of these scenarios could have significantly
  tilted the orbit relative to the host star's equatorial plane. Here
  we present spectroscopy of a transit of HD~147506b, and assess the
  spin-orbit alignment via the Rossiter-McLaughlin effect. We find the
  the sky projections of the stellar spin axis and orbital axis to be
  aligned within 14~deg. Thus we find no corroborating evidence for
  scattering or Kozai migration, although these scenarios cannot be
  ruled out with the present data.

\end{abstract}


\section{Introduction}

Giant planets that orbit Sun-like stars with periods shorter than
$\sim$5~days present both a problem and an opportunity. The problem is
how they achieved such tight orbits after presumably forming at much
larger orbital distances (Lin et al.~1996). The opportunity is that
such close-in planets are more likely to transit their parent stars,
giving access to many system properties such as the planetary radius,
temperature, and atmospheric composition, that are otherwise difficult
or impossible to measure (Charbonneau et al.~2006).

In this {\it Letter} we describe our attempt to exploit the transiting
configuration of the recently discovered exoplanet HD~147506b (Bakos
et al.~2007) to investigate the planet's particular migration
history. Interestingly, the orbit has a large eccentricity ($e\approx
0.5$), which is typical of giant planets at larger orbital distances,
but atypical of short-period planets. The expected $e$-folding time
for tidal circularization is comparable to the stellar age, making it
likely that some circularization has already occurred, and raising the
question of how such a high initial eccentricity was generated.

Simulations of inward planet migration via tidal interactions with the
protoplanetary disk generally do not predict eccentricities as large
as $0.5$ (see, e.g., D'Angelo, Lubow, \& Bate 2006).  In contrast,
planet-planet scattering naturally excites eccentricities to large
values (e.g., Rasio \& Ford 1996, Chatterjee et al.~2007).  Another
possibility is the Kozai mechanism: due to the tide of a third body,
the orbit undergoes eccentricity/inclination oscillations and
ultimately shrinks in semimajor axis due to tidal dissipation (e.g.,
Fabrycky \& Tremaine 2007, Wu et al.~2007). A corollary of either
scattering or Kozai migration is that the orbit can be tilted
considerably with respect to its initial orbital plane, which was
presumably close to the stellar equatorial plane.

One can search for such a misalignment by exploiting the
Rossiter-McLaughlin effect, the spectral distortion observed during a
transit due to stellar rotation. The planet hides part of the
rotational velocity field of the stellar photosphere, resulting in an
``anomalous Doppler shift'' (see, e.g., Ohta et al.~2005, Gim\'enez
2006, or Gaudi \& Winn~2007). The time sequence of anomalous Doppler
shifts depends on the angle between the stellar spin axis and the
orbital axis, as projected on the sky. This angle has been measured to
be small or consistent with zero in several systems, with accuracies
ranging from 1--30$\arcdeg$ (Bundy \& Marcy 2000; Queloz et al.~2000;
Winn et al.~2005, 2007a; Wolf et al.~2006; Narita et al.~2007).

Here we present observations of the RM effect for HD~147506, also
known as HAT-P-2. As reported by Bakos et al.~(2007), this system
consists of an F8 star with an unusually massive planet (8~$M_{\rm
  Jup}$) in a 5.6~day orbit. Our observations are described in \S~2,
our model in \S~3, and our results in \S~4, followed by a brief
summary and discussion.

\section{Observations}

We observed the transit of UT~2007~June~6 with the Keck~I 10m
telescope and the High Resolution Echelle Spectrometer (HIRES; Vogt et
al.~1994) following standard procedures of the California-Carnegie
planet search, as summarized here. We employed the red cross-disperser
and used the I$_2$ absorption cell to calibrate the instrumental
response and the wavelength scale. The slit width was $0\farcs85$ and
the typical exposure time was 200~s, giving a resolution of 70,000 and
a signal-to-noise ratio (SNR) of 200~pixel$^{-1}$. We obtained 97
spectra over 8.4~hr bracketing the predicted transit midpoint.

We determined the relative Doppler shifts with the algorithm of Butler
et al.~(1996). We estimated the measurement uncertainties based on the
scatter in the solutions for each 2~\AA~section of the spectrum. The
data are given in Table~1 and plotted in Fig.~1. Also shown in Fig.~1
are data obtained previously by Bakos et al.~(2007), consisting of 10
velocities measured with Keck/HIRES\footnote{Bakos et al.~(2007)
  reported 13 Keck/HIRES velocities measured on 10 different
  nights. For convenience, we binned the data that were obtained on
  the same night. We did not use the less precise Lick velocities, to
  avoid the introduction of another free parameter for the Lick/Keck
  velocity offset.} and a $z$-band transit light curve obtained with
the Fred L.~Whipple Observatory 1.2m telescope. All of these data were
incorporated into our model.
 
\begin{figure}[p]
\epsscale{1.0}
\plotone{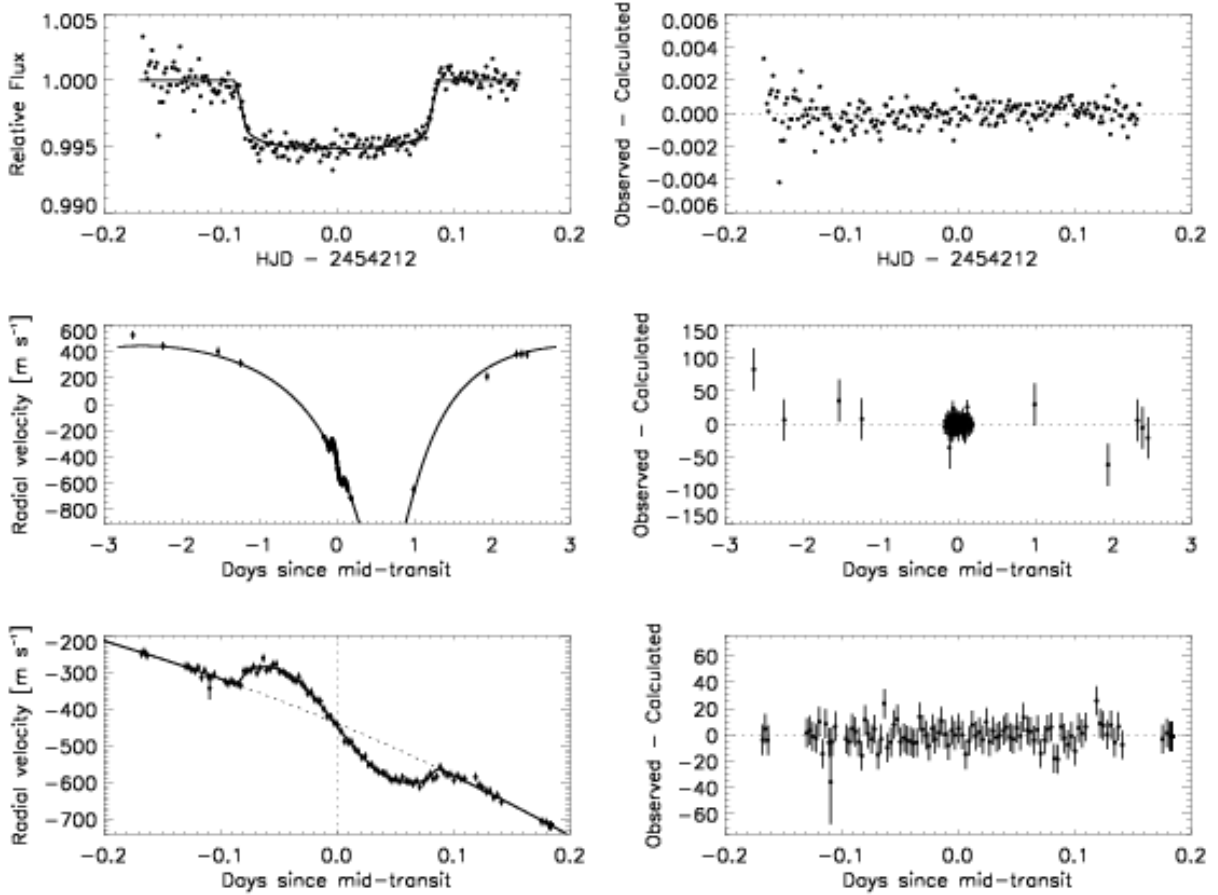}
\caption{
Photometry and spectroscopy of HD~147506.
{\it Top.} The $z$ band photometry of
Bakos et al.~(2007), binned into groups
of 4 points to reduce visual clutter.
The solid line is our best-fitting model.
{\it Middle.} Radial velocities,
from this work and from Bakos et al.~(2007),
as a function of the time modulo the orbital
period. The solid line is our best-fitting model.
{\it Bottom.} Close-up of the radial
velocities near the mid-transit time.
\label{fig:1}}
\end{figure}

\section{The Model}

We fitted the photometry and radial velocities with a parameterized
model based on a star and planet in a Keplerian orbit about the center
of mass. To calculate the relative flux as a function of the projected
separation of the planet and the star, we assumed the limb darkening
law to be quadratic and employed the formulas of Mandel \&
Agol~(2002). We fixed the $z$ band limb-darkening coefficients at the
values $a=0.14$, $b=0.36$, based on interpolation of the tables by
Claret~(2004) for a star with the observed effective temperature,
surface gravity, and metallicity.

To calculate the anomalous Doppler shift, we used the technique of
Winn et al.~(2005): we simulated RM spectra with the same data format
and noise characteristics as the actual data, and determined the
Doppler shifts using the same algorithm used on the actual data. The
simulations rely on a template spectrum (described below) that is
meant to mimic the emergent spectrum from a small portion of the
photosphere. We scaled the template spectrum in flux by $\epsilon$ and
shifted it in velocity by $v_p$, representing the spectrum of the
occulted portion of the stellar disk. We subtracted the scaled and
shifted spectrum from a rotationally-broadened version of the template
spectrum (broadened to 20~km~s$^{-1}$ to mimic the disk-integrated
spectrum of HD~147506), and then ``measured'' the anomalous Doppler
shift $\Delta v$. This was repeated for a grid of $\{\epsilon, v_p\}$,
and a polynomial function was fitted to the resulting surface.

The template spectrum should be similar to that of HD~147506 but with
narrower lines because of the lack of rotational broadening. We tried
both an empirical template and a theoretical template. The empirical
template was a spectrum of HD~3861 (F5, $v\sin
i_\star=2.8$~km~s$^{-1}$; Valenti \& Fischer 2005) that was observed
immediately following the transit, with a signal-to-noise ratio of 500
and a resolution of 70,000. The theoretical template had a resolution
of 200,000 and was taken from the model library of Coelho et
al.~(2005) for a nonrotating star with $T_{\rm eff}=6250$~K, $\log
g=4.0$, and [Fe/H]~$=0.10$. The final results for the system
parameters did not vary significantly with the choice of template, but
the calibration based on the empirical template provided a better fit
to the data, and hence in our final results we used the following
relation based on the empirical template:
\begin{equation}
\Delta v =
-\epsilon~v_p
\left[
  2.16
- 2.47 \left( \frac{v_p}{{\rm 20~km~s}^{-1}} \right)^2
+ 0.98 \left( \frac{v_p}{{\rm 20~km~s}^{-1}} \right)^4
\right].
\end{equation}
With this formula, the anomalous Doppler shift $\Delta v(t)$ can be
related to the flux decrement $\epsilon$ and the sub-planet velocity
$v_p$ at that time. The sub-planet velocity is the projected rotation
velocity of the portion of the star hidden by the planet. When fitting
the model to the data, $v_p$ was computed as a function of the
relative position of the star and planet under the assumption of
uniform rotation of the photosphere.

The fitting statistic was
\begin{eqnarray}
\chi^2 & = &
\sum_{j=1}^{962}
\left[
\frac{f_j({\mathrm{obs}}) - f_j({\mathrm{calc}})}{\sigma_{f,j}}
\right]^2
+
\sum_{j=1}^{111}
\left[
\frac{v_j({\mathrm{obs}}) - v_j({\mathrm{calc}})}{\sigma_{v,j}}
\right]^2
+
\left( \frac{v\sin i_\star - 19.8~{\rm km~s}^{-1}}{1.6~{\rm km~s}^{-1}} \right)^2
+ \nonumber \\
& & 
\left( \frac{M_\star/M_\odot - 1.32}{0.08} \right)^2
+
\left( \frac{R_\star/R_\odot - 1.53}{0.10} \right)^2
+
\left( \frac{\Delta\gamma}{{\rm 31~m~s^{-1}}} \right)^2
,
\end{eqnarray}
where $f_j$(obs) is the flux observed at time $j$, $\sigma_{f,j}$ is
the corresponding uncertainty, and $f_j$(calc) is the calculated
value. A similar notation applies to the velocities. The last four
terms are a priori constraints. The first three of these constraints
enforce the spectroscopic parameters derived by Bakos et
al.~(2007). The last constraint is explained below.

It is important for the data weights $\sigma_{f,j}$ and $\sigma_{v,j}$
to account for unmodeled systematic errors in addition to measurement
errors. We assessed systematic errors in the photometry by examining
the out-of-transit (OOT) measurements. We compared the standard
deviation of the unbinned OOT data ($\sigma_1$) with the standard
deviation of the binned OOT data ($\sigma_N$, where $N=40$ data points
or 20 minutes). We chose the bin size to be the approximate ingress or
egress duration, since those phases provide much of the leverage on
the model parameters. With independent Gaussian noise, we would find
$\sigma_N=\sigma_1/\sqrt{40}$ but in fact $\sigma_N$ was larger by a
factor of 1.25, representing a level of ``red noise'' that is commonly
seen in ground-based photometry. For this reason, we set $\sigma_{f,j}
= 1.25~\sigma_1$ before computing our final results. With this choice,
the minimum value of $\chi^2/N_{\rm DOF}$ was less than unity, which
is appropriate because the data points are not truly independent.

Likewise, the weights for the velocities should account for intrinsic
velocity noise (``photospheric jitter''). We assessed the amplitude
and time scale of the noise as follows. First, we fitted only the 10
velocities obtained by Bakos et al.~(2007) on different nights. We
fixed the orbital period and transit time and optimized the velocity
semiamplitude, eccentricity, argument of pericenter, and velocity zero
point. The root-mean-squared (RMS) residual was 32~m~s$^{-1}$. This is
consistent with the quadrature sum of the typical measurement error of
7~m~s$^{-1}$ and an intrinsic noise term of 31~m~s$^{-1}$, and the
latter is in agreement with the empirical noise estimators of
Wright~(2005) for an F8 star with the observed rotation
velocity. Therefore, for fitting purposes, we took the weights
$\sigma_{v,j}$ of the Bakos et al.~(2007) velocities to be the
quadrature sum of the measurement error and 31~km~s$^{-1}$. Second, we
fitted only 31 OOT velocities observed on 2007~Jun~6, finding the RMS
residual to be 11~m~s$^{-1}$, which is consistent with the quadrature
sum of the measurement error and an intrinsic noise term of
10~m~s$^{-1}$. Therefore, for the data taken on 2007~Jun~6, we
calculated $\sigma_{v,j}$ by adding the measurement error and
10~m~s$^{-1}$ in quadrature.  Apparently, most of the intrinsic
velocity noise occurs on a time scale longer than one night, as we
also found for HD~189733 (Winn et al.~2006).

The model parameters were the two bodies' masses and radii ($M_\star$,
$M_p$, $R_\star$ and $R_p$); the orbital inclination ($i$); the
mid-transit time ($T_c$); the line-of-sight stellar rotation velocity
($v \sin i_\star$); the angle between the projected stellar spin axis
and orbit normal ($\lambda$); the velocity zero point ($\gamma$); and
a velocity offset specific to the night of 2007~Jun~6
($\Delta\gamma$). This last parameter is needed because of the
photospheric jitter; the last term in Eq.~(2) enforces a reasonable
level of noise. We fixed the orbital period to be $5.63341$~days
(Bakos et al.~2007). We used a Markov Chain Monte Carlo algorithm to
solve for the model parameters and their 68\% confidence
limits.\footnote{For the background on this method, see Tegmark et
  al.~(2004), Ford (2005) or Burke et al.~(2007); for more detail on
  our particular implementation, see, Holman et al.~(2006) or Winn et
  al.~(2007).}

\section{Results}

The results are given in Table~2. Our results for both $R_p$ and
$R_\star$ are in agreement with the values determined by Bakos et
al.~(2007), which is not surprising, given that those parameters
depend chiefly on the photometry and we have used precisely the same
photometry.\footnote{To forestall possible confusion, we note that an
  early version of the manuscript by Bakos et al.~(2007) that was
  distributed on arxiv.org quoted significantly larger values for
  $R_p$ and $R_\star$. This is because those authors had not yet made
  use of the measured transit duration in determining the system
  parameters, as their original model did not take into account the
  nonuniform speed of the planet.} The impact parameter $b$ of a
transit is the minimum projected star-planet distance, expressed in
units of the stellar radius. For an eccentric orbit it is given
approximately by
\begin{equation}
b = \frac{1-e^2}{1+e\sin\omega}~\frac{a\cos i}{R_\star}.
\end{equation}
We find that the HD~147506 transit occurs at a small impact parameter:
$b<0.41$ with 95\% confidence. This follows from the short durations
of ingress and egress relative to the total duration of the transit.
This upper limit on the impact parameter is more constraining than the
upper limit that was obtained by Bakos et al.~(2007), indicating that
our transit velocities are providing most of the leverage on the
impact parameter.

As for the key spin-orbit parameter, we found $\lambda = 1.2\pm
13.4$~deg, i.e., consistent with perfect alignment. The 95\%
confidence upper limit on $|\lambda|$ is 29.6\arcdeg. There is a
strong covariance between $\lambda$ and $v\sin i_\star$, which is a
consequence of the small impact parameter (Gaudi \& Winn~2007). For
this reason, we investigated the dependence of our results on the a
priori constraint on $v\sin i_\star$, by either abandoning the
constraint or by strengthening it. The conclusion that $\lambda$ is
consistent with zero is unchanged by modifying the constraint; only
the dispersion in $\lambda$ changes. If we drop the constraint
completely, the 1~$\sigma$ error in $\lambda$ grows to 23\arcdeg, and
the result for $v\sin i_\star$ is $20\pm 3$~km~s$^{-1}$. However, if
we assume $v\sin i_\star=20$~km~s$^{-1}$ exactly, then the error in
$\lambda$ shrinks to 9\arcdeg.

\section{Summary and Discussion}

We have monitored the apparent Doppler shift of HD~147506 during a
transit of its giant planet, and have modeled the available
photometric and spectroscopic data. By modeling the RM effect, we find
that the stellar spin axis and the orbit normal are aligned on the sky
to within $14\arcdeg$. This is unlikely to be a coincidence. If the
spin axis were randomly oriented, the probability of observing
$|\lambda|$ smaller than $14\arcdeg$ would be $\approx 14/180 =
7.7\%$.  It is also reasonable to suppose that the other component of
the stellar spin vector (that which is directed towards or away from
the Earth) is of the same order of magnitude as $\lambda$, because our
viewing perspective is random and because the observed value of $v\sin
i_\star$ is typical of a main-sequence F8 star (Cox et al.~2000,
p.~389).

To interpret this result, we first calculate the expected timescale
for tidal interactions to cause the stellar obliquity to decay, within
the framework of Hut's~(1981) analytic model of equilibrium tides.
Assuming a stellar tidal quality factor $Q'\sim10^6$, we
find\footnote{This is estimated from Eqs.~(12), (22), (49), and (53)
  of Hut~(1981).} $\tau \sim 10^{10}$~yr, which is longer than the
estimated stellar age of $3\times 10^9$~yr (Bakos et al.~2007). This
suggests that the alignment we observe today was the outcome of the
planet migration process, rather than an aftereffect of tidal
interactions.\footnote{Likewise, the stellar spin is apparently not
  yet synchronized with the orbit. Assuming $\sin i_\star =1$, the
  stellar rotation period is $2\pi R_\star/v= 3.8$~days, which is in
  between the orbital period of 5.6~days and the pseudosynchronization
  period of 1.9~days for $e=0.5$ [see Eq.~(42) of Hut~(1981)].} Thus,
HD~147506b is in the same category of well-aligned planetary orbits as
the other measured systems, despite its uniquely eccentric orbit.

Had the outcome of this experiment been a significant misalignment, it
would have argued against quiescent migration due to tidal interaction
with a protoplanetary disk, and invited an interpretation as either
the outcome of a planet-planet scattering event or Kozai oscillations
accompanied by tidal dissipation. However, the actual outcome does not
rule out the latter scenarios. Planet-planet scattering does enhance
any initial misalignments between an initial planetary orbit and other
orbits, as well as the stellar spin axis. Indeed, Chatterjee, Ford, \&
Rasio~(2007) have predicted a broad range of inclinations ranging up
to $45\arcdeg$ for hot Jupiters produced in this manner. However, it
is possible that the particular impulse that threw HD~147506b inwards
had only a small vertical component. The same simulations by
Chatterjee et al.~(2007) found that about half of the planets
scattered inward experience misalignments smaller than
$15\arcdeg$. Likewise, Fabrycky \& Tremaine~(2007) have predicted the
distribution of stellar obliquities that should result from Kozai
migration due to isotropically placed stellar companions. There is a
broad distribution ranging out to 140\arcdeg, but the final obliquity
is smaller than 20\arcdeg\ approximately 20\% of the
time.\footnote{The Kozai scenario may be more strongly disfavored than
  this simple comparison suggests, because those systems with small
  final obliquities also tended to have final eccentricities much
  smaller than the observed eccentricity of 0.5 (D.~Fabrycky 2007,
  private communication).} It should be noted that neither Chatterjee
et al.~(2007) nor Fabrycky \& Tremaine~(2007) intended their
calculations for direct comparison for observations, but such
calculations now seem warranted, given the increasing number of
accurate Rossiter-McLaughlin measurements such as the one presented
here.

In this case, the accuracy with which $\lambda$ can be measured is
hampered by the small impact parameter of the transit. Further
improvement is possible if the impact parameter can be measured with
greater precision and shown to be inconsistent with zero. However,
transits with impact parameters near 0.5 offer much greater
sensitivity to $\lambda$ and a cleaner separation from $v\sin i_\star$
(Gaudi \& Winn 2007). For this reason, the discovery of additional
transiting eccentric planets are eagerly anticipated, and seem bound
to happen soon, given the rapidity with which new transiting systems
are being announced.

\acknowledgments We thank D.~Fabrycky and S.~Gaudi for helpful
comments on the manuscript. We are very grateful to Michael Bolte,
Jennifer Johnson, and David Lai for swapping Keck nights on short
notice. G.\'{A}.B.~was supported by NASA through a Hubble Fellowship
Grant HST-HF-01170.01. We recognize and acknowledge the very
significant cultural role and reverence that the summit of Mauna Kea
has always had within the indigenous Hawaiian community. We are most
fortunate to have the opportunity to conduct observations from this
mountain.

\begin{deluxetable}{lcccc}
\tabletypesize{\normalsize}
\tablecaption{System Parameters of HD~147506\label{tbl:params}}
\tablewidth{0pt}

\tablehead{
\colhead{Parameter} & \colhead{Value} & \colhead{Uncertainty}
}

\startdata
                                           $M_\star/M_\odot$& $          1.32$  & $           0.08$ \\
                                           $M_p/M_{\rm Jup}$& $          8.04$  & $           0.40$ \\
                                           $R_\star/R_\odot$& $          1.48$  & $           0.05$ \\
                                           $R_p/R_{\rm Jup}$& $          0.98$  & $           0.04$ \\
                     Orbital period~[days]\tablenotemark{a} & $       5.63341$  & $        0.00013$ \\
                                                         $e$& $         0.501$  & $          0.007$ \\
                                              $\omega$~[deg]& $        -172.6$  & $            1.6$ \\
                                                   $i$~[deg]& $         >86.8$  & (95\% conf.) \\
                          Impact parameter\tablenotemark{b} & $         <0.41$  & (95\% conf.) \\
                                                 $T_c$~[HJD]& $  2454212.8561$  & $         0.0006$ \\
                              $v \sin i_\star$~[km~s$^{-1}$]& $          19.6$  & $            1.0$ \\
                                             $\lambda$~[deg]& $           1.2$  & $           13.4$ \\
                                       $\gamma$~[m~s$^{-1}$]& $           399$  & $             10$ \\
                                 $\Delta\gamma$~[m~s$^{-1}$]& $            45$  & $             18$
\enddata

\tablenotetext{a}{From Bakos et al.~(2007).}

\tablenotetext{b}{See Eq.~(3).}

\end{deluxetable}

\end{document}